\documentclass[twocolumn,superscriptaddress,preprintnumbers,amsmath,amssymb]{revtex4}
\usepackage{graphicx}
\usepackage{dcolumn}
\usepackage{bm}
\usepackage{hyperref}
\usepackage{amssymb,amsmath,amsthm,amsfonts}
\usepackage{bbm}
\usepackage{mathtools}
\usepackage{color}
\usepackage{textcomp}
\usepackage{gensymb} 
\usepackage{float}
\usepackage{soul}
\usepackage{booktabs}
\usepackage{graphicx}
\usepackage{siunitx}
\usepackage{multirow}
\usepackage{makecell}
\usepackage{draftwatermark}

\input{Qcircuit}              

\newcommand{\eq}[1]{(\ref{eq:#1})}
\renewcommand{\sec}[1]{\hyperref[sec:#1]{Section~\ref*{sec:#1}}}
\newcommand{\fig}[1]{\hyperref[fig:#1]{Figure~\ref*{fig:#1}}}
\newcommand{\tab}[1]{\hyperref[tab:#1]{Table~\ref*{tab:#1}}}

\global\long\def\l({\left(}
\global\long\def\r){\right)}

\newcommand{\Gate}[1]{\textsc{#1}}
\newcommand{\xxgate}{\Gate{XX}}
\newcommand{\xxcirc}{\Gate{xx}}
\newcommand{\hgate}{\Gate{H}}
\newcommand{\hcirc}{\Gate{h}}

\newcommand{\sgate}{\Gate{S}}
\newcommand{\scirc}{\Gate{s}}

\newcommand{\notgate}{\Gate{NOT}}
\newcommand{\cnotgate}{\Gate{CNOT}}
\newcommand{\swapgate}{\Gate{SWAP}}

\newcommand{\rzcirc}{{\Gate{r}}_{z}}

\newcommand{\comment}[1]{\textbf{\color{red}}}

\makeatletter
\newcommand{\vast}{\bBigg@{4}}
\newcommand{\Vast}{\bBigg@{5}}
\newcommand{\VVast}{\bBigg@{11.8}}
\makeatother

\begin{document}


\title{Ground-state energy estimation of the water molecule on a trapped ion quantum computer}

\author{Yunseong Nam}
\email{nam@ionq.co}
\affiliation{IonQ, Inc., College Park, MD 20740, USA}
\author{Jwo-Sy Chen}
\affiliation{IonQ, Inc., College Park, MD 20740, USA}
\author{Neal C. Pisenti}
\affiliation{IonQ, Inc., College Park, MD 20740, USA}
\author{Kenneth Wright} 
\affiliation{IonQ, Inc., College Park, MD 20740, USA}
\author{Conor Delaney}
\affiliation{IonQ, Inc., College Park, MD 20740, USA}
\author{Dmitri~Maslov}  
\affiliation{National Science Foundation, Alexandria, VA 22314, USA}
\author{Kenneth R. Brown} 
\affiliation{IonQ, Inc., College Park, MD 20740, USA}
\affiliation{Department of Electrical and Computer Engineering, Duke University, Durham, NC 27708, USA}
\author{Stewart Allen}
\affiliation{IonQ, Inc., College Park, MD 20740, USA}
\author{Jason M. Amini}
\affiliation{IonQ, Inc., College Park, MD 20740, USA}
\author{Joel Apisdorf}
\affiliation{IonQ, Inc., College Park, MD 20740, USA}
\author{Kristin M. Beck}
\affiliation{IonQ, Inc., College Park, MD 20740, USA}
\author{Aleksey Blinov} 
\affiliation{IonQ, Inc., College Park, MD 20740, USA}
\author{Vandiver Chaplin} 
\affiliation{IonQ, Inc., College Park, MD 20740, USA}
\author{Mika Chmielewski}  
\affiliation{IonQ, Inc., College Park, MD 20740, USA}
\affiliation{Joint Quantum Institute and Department of Physics, University of Maryland, College Park, MD 20742, USA}
\author{Coleman Collins} 
\affiliation{IonQ, Inc., College Park, MD 20740, USA}
\author{Shantanu Debnath} 
\affiliation{IonQ, Inc., College Park, MD 20740, USA}
\author{Andrew M. Ducore}
\affiliation{IonQ, Inc., College Park, MD 20740, USA}
\author{Kai M. Hudek}
\affiliation{IonQ, Inc., College Park, MD 20740, USA}
\author{Matthew Keesan}
\affiliation{IonQ, Inc., College Park, MD 20740, USA}
\author{Sarah M. Kreikemeier}
\affiliation{IonQ, Inc., College Park, MD 20740, USA}
\author{Jonathan Mizrahi} 
\affiliation{IonQ, Inc., College Park, MD 20740, USA}
\author{Phil Solomon}
\affiliation{IonQ, Inc., College Park, MD 20740, USA}
\author{Mike Williams}
\affiliation{IonQ, Inc., College Park, MD 20740, USA}
\author{Jaime David Wong-Campos}
\affiliation{IonQ, Inc., College Park, MD 20740, USA}
\author{Christopher Monroe}
\affiliation{IonQ, Inc., College Park, MD 20740, USA}
\affiliation{Joint Quantum Institute and Department of Physics, University of Maryland, College Park, MD 20742, USA}
\author{Jungsang~Kim}
\email{kim@ionq.co}
\affiliation{IonQ, Inc., College Park, MD 20740, USA}
\affiliation{Department of Electrical and Computer Engineering, Duke University, Durham, NC 27708, USA}


\begin{abstract}
Quantum computing leverages the quantum resources of superposition and entanglement to efficiently solve computational problems considered intractable for classical computers.
Examples include calculating molecular and nuclear structure, simulating strongly-interacting electron systems, and modeling aspects of material function.
While substantial theoretical advances have been made in mapping these problems to quantum algorithms, there remains a large gap between the resource requirements for solving such problems and the capabilities of currently available quantum hardware. 
Bridging this gap will require a co-design approach, where the expression of algorithms is developed in conjunction with the hardware itself to optimize execution.
Here, we describe a scalable co-design framework for solving chemistry problems on a trapped ion quantum computer, and apply it to compute the ground-state energy of the water molecule.
The robust operation of the trapped ion quantum computer yields energy estimates with errors approaching the chemical accuracy, which is the target threshold necessary for predicting the rates of chemical reaction dynamics.
\end{abstract}

\maketitle

Quantum computation has attracted much attention for its potential to solve certain computational problems that are difficult to tackle with classical computers.
For example, integer factorization\cite{ar:Shor}, unsorted database search\cite{ar:Grover}, and the simulation of quantum systems\cite{ar:Feynman} admit quantum algorithms that outperform the best-known classical algorithms given a sufficiently large problem size.
However, these algorithms require substantial quantum resources to achieve a practical advantage over classical techniques, limiting their near-term utility on noisy intermediate-scale quantum (NISQ) devices\cite{ar:NISQ} that are severely limited in the number of gates they can perform before errors dominate the output.
Any useful quantum computation on a NISQ device will require further advances in hardware performance, as well as advances in algorithmic design.

Quantum chemistry is a promising application where quantum computing might overcome the limitations of known classical algorithms, hampered by an exponential scaling of computational resource requirements. One of the most challenging tasks in quantum chemistry is to determine molecular energies to within  chemical accuracy, defined to be the target accuracy necessary to estimate chemical reaction rates at room temperature and generally taken to be $\approx \SI[per-mode=symbol]{4}{\kilo\joule\per\mole} = \num{1.6e-3}$~Hartree (Ha)\cite{Yuan2018}.
Achieving chemical accuracy would allow computational methods to replace costly experimental procedures in chemical and materials engineering, augmenting these fields to accelerate the pace of discovery.

Early quantum computational techniques to simulate many-body Fermi systems\cite{PhysRevLett.79.2586} or calculate molecular energies\cite{Aspuru-Guzik1704} 
have dramatically improved over the past decade\cite{HastingsQIC2015,BabbushNJP2016,BerryNQI2018}, but the resource requirements for useful chemical simulations still remain out of reach\cite{Reiher7555}.
Hybrid approaches might relax these requirements, where a short quantum computation serves as a subroutine to calculate classically difficult quantities.  
The variational quantum eigensolver (VQE) method is one example, which estimates the ground state of a system by positing an ansatz state defined by a set of variational parameters and minimizing its energy.
The quantum subroutine determines the energy for a particular set of ansatz parameters, and a classical optimization algorithm iteratively updates the ansatz to reduce the energy until it converges.
Early demonstrations of the VQE method have been performed on different quantum architectures\cite{ar:VQE,ar:ChemDemon3,ar:ChemDemon1,ar:ChemDemon2}, but they rely on dramatic simplifications that work for small molecules but do not readily generalize to larger systems.
Additionally, systematic errors in the experimental results far exceed chemical accuracy due to imperfections in the quantum computer (QC) hardware.

Here, we provide a generic VQE approach that scales to much larger molecular systems and use it to compute the ground-state energy of the water molecule ($\rm H_2\rm O$).
We embrace co-design principles to fully optimize the quantum circuits for a trapped-ion QC, and experimentally compute the first three correction terms beyond the mean-field (Hatree-Fock) approximation.  We achieve computational errors approaching chemical accuracy for the first time, without using any error mitigation techniques.
These results establish a path for future computations on more complex systems as trapped-ion QCs continue to improve, eventually reaching beyond the capability of classical methods.

\section{Trapped-ion quantum computer}
\label{sec:Hardware}

\begin{figure*}[t]
\centering
\includegraphics[width=\textwidth]{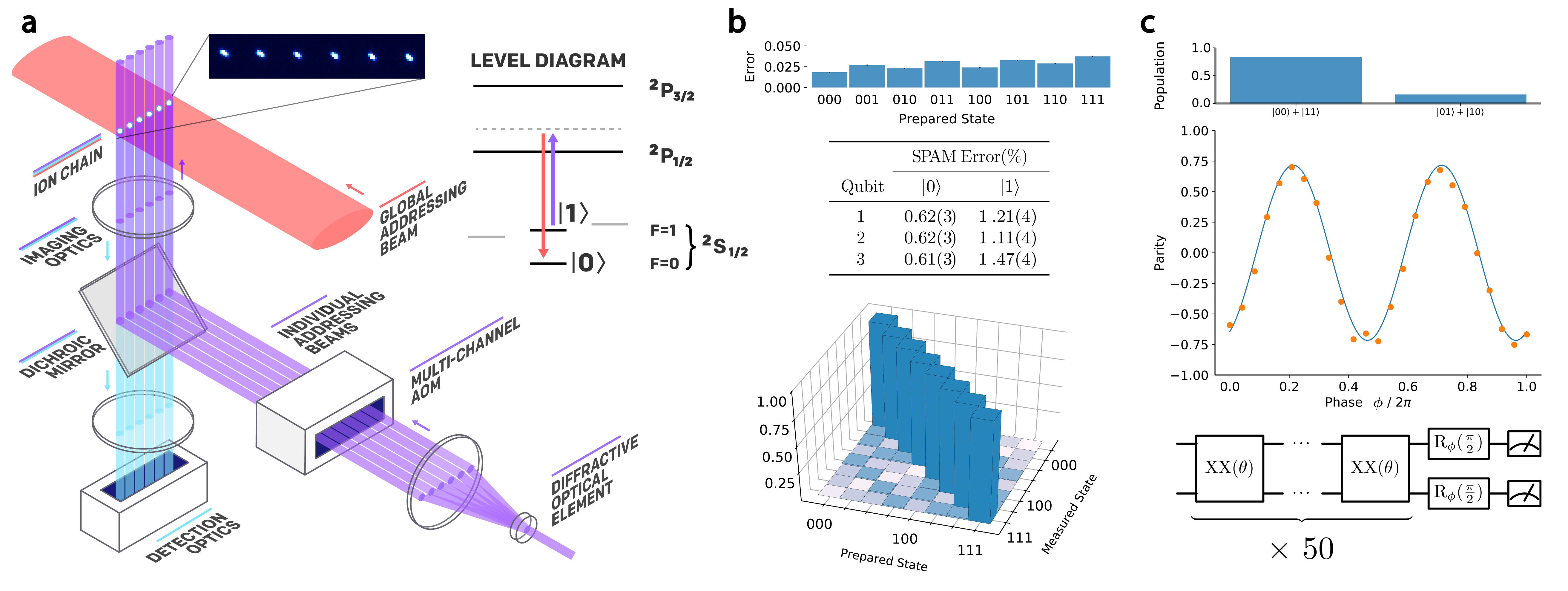} 
\caption{\textbf{Apparatus and performance.} \textbf{a}, Schematic representation of the trapped-ion QC.  The qubit register is implemented in a linear chain of $^{171}\text{Yb}^+$ ions residing inside an ultra-high vacuum chamber (not shown), and high-NA imaging optics enable individual addressing and readout of the ion qubits.
The Raman beams (shown in red and purple) are generated from a pulsed laser at \SI{355}{\nano\meter} and drive a two-photon transition between $|0\rangle$ and $|1\rangle$. Full control of the amplitude, frequency, and phase of the individual addressing beams enables implementations of arbitrary single-- and two--qubit operators.
\textbf{b}, SPAM characterization on a three-ion chain. From top to bottom, we show the SPAM error of each three-qubit state, the per-qubit SPAM error for $|0\rangle$ and $|1\rangle$, and a bar plot of the full SPAM matrix where the color is log-scaled for visibility. We see no indication of measurement crosstalk between qubits.
\textbf{c}, Characterization of the small-angle $\xxgate(\theta)$~gate performance. The state fidelity after 50 consecutive small-angle \xxgate~gates is $\approx 78\%$, and we estimate the per-gate error to be $\epsilon \lesssim \num{4e-3}$.
}
\label{fig:apparatus}
\end{figure*}

The trapped-ion system used in this study is a scalable, general-purpose programmable QC constructed at IonQ, Inc.\cite{www:IonQ} and illustrated schematically in Fig.~\ref{fig:apparatus}a; see Methods for additional details.
The computer consists of a linear chain of $^{171}\text{Yb}^+$ ions on a surface trap operating at room temperature, where the qubit is implemented between the ${|0\rangle \equiv |F=0\text{,}\,m_F=0\rangle}$ and the ${|1\rangle \equiv |F=1\text{,}\,m_F=0\rangle}$ hyperfine levels of the $^{2}S_{1/2}$ ground state of each ion, split by \SI{12.6}{\giga\hertz}\cite{Yb:2007}.

The qubit register is initialized to the $|0\rangle$ state using optical pumping, and measured at the end of the computation by state-dependent fluorescence on the dipole-allowed cycling transition between ${| 1 \rangle}$ and the $^{2}P_{1/2}$ excited state\cite{Noek:13}. 
Scattered photons from the ions during detection are collected through a high numerical aperture lens ($\text{NA} \approx 0.6$) and passed through a dichroic mirror to an array of photon detectors for simultaneous readout of the entire qubit register. 
State preparation and measurement (SPAM) errors are routinely characterized during computation, with typical data for a three-ion chain shown in Fig.~\ref{fig:apparatus}b. Our system exhibits a small asymmetry in the SPAM error for $|0\rangle$ versus $|1\rangle$ (0.6\% and 1.3\%, respectively), which is well understood from an atomic model of the detection process\cite{acton2006near}. 
We observe no evidence of measurement crosstalk where the state of one qubit affects the readout of neighboring qubits, which allows SPAM correction to be performed with low overhead. 
SPAM errors can be readily improved by increasing the collection efficiency of the detection optics\cite{acton2006near,MyersonPRL2008,Noek:13}.

Quantum gates are implemented via two-photon Raman transitions driven by two laser beams from a mode-locked pulsed laser at \SI{355}{\nano\meter}, where the two laser beams generate a beatnote close to the qubit  frequency\cite{PhysRevLett.104.140501}.
One of the beams is a ``global'' beam, with a wide profile that uniformly illuminates all qubits in the chain.
The other is an array of tightly focused beams, generated from a diffractive optical element and a multi-channel acousto-optic modulator (AOM), that address the ions individually.
By controlling the phase, frequency, and amplitude of these beams, we can manipulate individual qubits to implement arbitrary quantum logic gates\cite{DebnathNature2016}. 
The AOM in our system has 32 independent channels, allowing us to scale the number of individually addressable and fully-connected qubits to this number.  
Further scaling is possible with alternative optical setups or by sacrificing full connectivity and using ion-shuttling protocols\cite{Kielpinski2002}.

We drive high-fidelity single-qubit operations with a resonant Raman transition between $|0\rangle$ and $|1\rangle$ using a composite pulse sequence\cite{Brown:2004,PhysRevA.92.060301}. 
Two-qubit operations are mediated by the shared motional modes of the entire chain via an effective $\xxgate$-Ising interaction using the {M\o lmer}-{S\o rensen} protocol\cite{PhysRevLett.82.1971,Choi:XX}, and can be written in terms of Pauli X matrices on ions $i$ and $j$ as $\xxgate(\theta) = \exp\left[-i\theta \sigma_x^i\sigma_x^j/2\right]$.
Since the motional modes involve every ion in the chain, we can apply the \xxgate~gate between arbitrary pairs of ions with comparable speed and fidelity\cite{ZhuEL2006,ZhuPRL2006,DebnathNature2016,ar:ATA}. 
This native all-to-all connectivity of  two-qubit gates in the trapped ion QC provides complete flexibility to choose qubit mappings and gate configurations that maximize  circuit performance on the hardware. 
Under typical operating conditions for this QC, the single-qubit gate fidelity can be maintained $\gtrsim 99.9\%$, and the state fidelity of a maximally entangling $\xxgate(\pi/2)$ gate is $\gtrsim 96\%$.
We estimate the fidelity of ``small-angle'' \xxgate~gates by concatenating  $\xxgate(\pi/2n)$ gates $n$ times to approximate a full $\xxgate(\pi/2)$~gate.
The state fidelity $\mathcal F$ is measured, and we estimate the per-gate error to be $\epsilon \lesssim (1 - \mathcal F)/n$.
We show an example in Fig.~\ref{fig:apparatus}c for $n=50$, and calculate $\epsilon\lesssim\num{4e-3}$ for the $\xxgate(\pi/100)$ gate.
In general, small-angle \xxgate~gates have higher fidelities in a trapped-ion QC than maximally entangling $\xxgate$ gates, which can be used to improve the quality of a quantum computation.

\section{Molecular modeling}
\label{sec:Results}

\begin{figure*}[ht]
    \centering
    \includegraphics[width=0.9\textwidth]{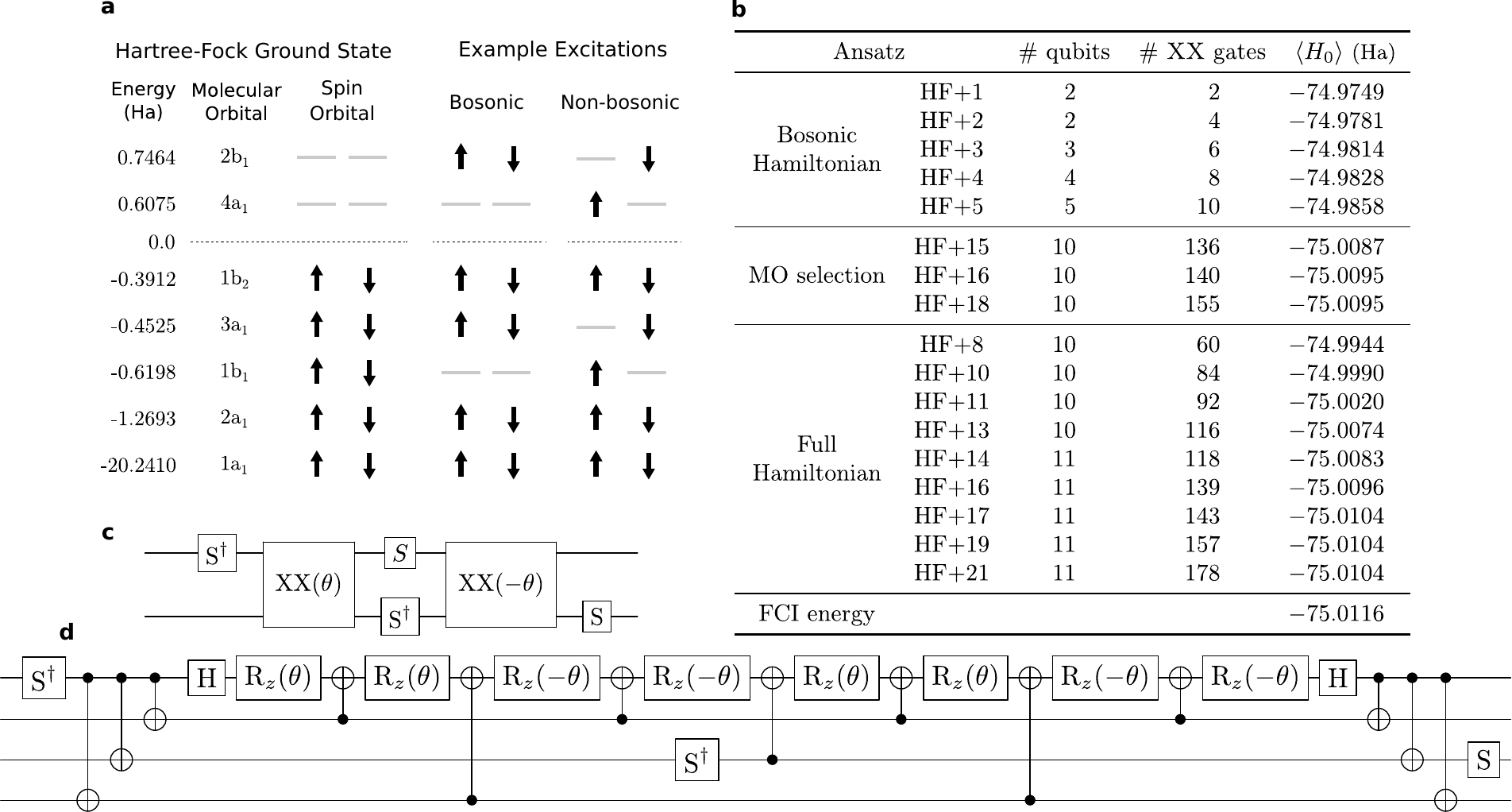}
    \caption{\textbf{Circuit design for quantum chemistry.} 
    \textbf{a}, Molecular and spin orbital diagram, and examples of bosonic and non-bosonic excitations from the Hartree-Fock (HF) ground state.
    \textbf{b}, Metrics for each circuit, labeled HF+$N$, as up to $N$ of the most significant interaction terms are added to the ansatz state.
    The bosonic terms through HF+5 can be represented as pair excitations to reduce the qubit resource requirements, while the MO selection strategy prunes the two least significant molecular states ($1a_1$ and $1b_2$) to reduce the qubit count slightly at the expense of $\approx$~mHa accuracy.
    Energies should be compared to the (FCI) ground state energy, which is the exact result from diagonalizing the complete Hamiltonian in the minimal STO-3G chemical basis.
    \textbf{c}, Bosonic excitation template circuit.
    \textbf{d}, Non-bosonic excitation template circuit.}
    \label{tab:circ_metric}
\end{figure*}

We choose ${\rm H}_2 {\rm O}$ as a testbed for  quantum co-design principles.
The structure of ${\rm H}_2 {\rm O}$ is sufficiently complex to develop and test universal techniques for scalable quantum circuit synthesis, while simple enough to be accessible by currently available trapped-ion QCs.
Simulations using classical hardware provide fully verified solutions to assess the performance of the quantum hardware, and build intuition about successful co-design strategies.
What follows is a brief summary of the VQE co-design methodology, with further details supplied in the Methods.

We first write down a Hamiltonian under the Born-Oppenheimer approximation, where the atomic nuclei are fixed to their known equilibrium geometry.
The Hamiltonian is represented in the second-quantized form
\begin{equation} 
\label{eq:hamiltonian}
\hat{H} = \sum_{p,q}{h_{pq} c_p^\dagger c_q} + \sum_{p,q,r,s}{h_{pqrs} c_p^\dagger c_q^\dagger c_r c_s}\,,
\end{equation}
where $c^\dagger_p$~($c_p$) are the creation (annihilation) operators for a molecular spin-orbital (SO) $p$. The SOs are spin-labeled molecular orbitals (MOs) obtained as a linear combination of atomic orbitals from the minimal {STO-3G} chemical basis\cite{ar:STO3G} using the Hartree-Fock (HF) method\cite{bk:MC}. The resulting 7 MOs (14 SOs) are shown schematically in Figure~\ref{tab:circ_metric}a,
and the terms $h_{pq}$ and $h_{pqrs}$ from equation~(\ref{eq:hamiltonian}) are computed classically using a standard open-source tool based on \textit{ab initio} methods\cite{ar:psi4}. 
The $c_p$  and $c^\dagger_p$ operators can be represented as Pauli operators acting on individual qubits using the Jordan-Wigner (JW) transformation\cite{ar:JW},
and we use the unitary coupled-cluster (UCC) method to generate an ansatz state\cite{ar:CC,ar:UCC1,ar:UCC2} with the first-order Trotter formula and one Trotter step.
The expectation value of the Hamiltonian is computed by measuring projections of the prepared ansatz state in the combination of Pauli bases that correspond to each term in the JW-transformed Hamiltonian.
To achieve meaningful accuracy, the circuit must be sufficiently sampled in each basis to reduce statistical errors\cite{ar:ChemDemon1}, and systematic errors must be controlled.

For a small molecule like ${\rm H}_2 {\rm O}$ in the minimal basis set, it is possible to diagonalize the Hamiltonian in equation~(\ref{eq:hamiltonian}) to compute the full configuration-interaction (FCI) ground state energy ($-75.0116$~Ha).
This energy is lower than the mean-field HF result ($-74.9624$~Ha) by ${\sim} 49.2$~mHa. 
From the FCI diagonalization, we generate a list of two-electron interaction terms ($c_p^\dagger c_q^\dagger c_r c_s$) that contribute to modifications in the energy during the diagonalization process, with the degree of contribution characterized by the determinant. 
Some of these terms correspond to a pair of spin-up and spin-down electrons from the same filled MO being simultaneously excited to an empty MO (called the ``bosonic'' excitation terms hereafter), and the rest correspond to excitations of two electrons that are not paired in this way (see examples in Fig.~\ref{tab:circ_metric}a).
Each term can be included in the preparation of the UCC ansatz in the form of $\exp\left[\theta_{pqrs}c_p^\dagger c_q^\dagger c_r c_s-c.c.\right]$ in the Trotter product formula, where $\theta_{pqrs}$ becomes the optimization parameter and $h.c.$ denotes the Hermitian conjugate operator.
We perform a numerical simulation of the VQE process as more terms are added to the UCC ansatz, and estimate the lowest energy for each ansatz state as the parameters are optimized. 
This \textit{in-silico} result serves as a reference to benchmark the computational outcome from the QC.

\begin{figure*}
    \centering
    \includegraphics[width=\textwidth]{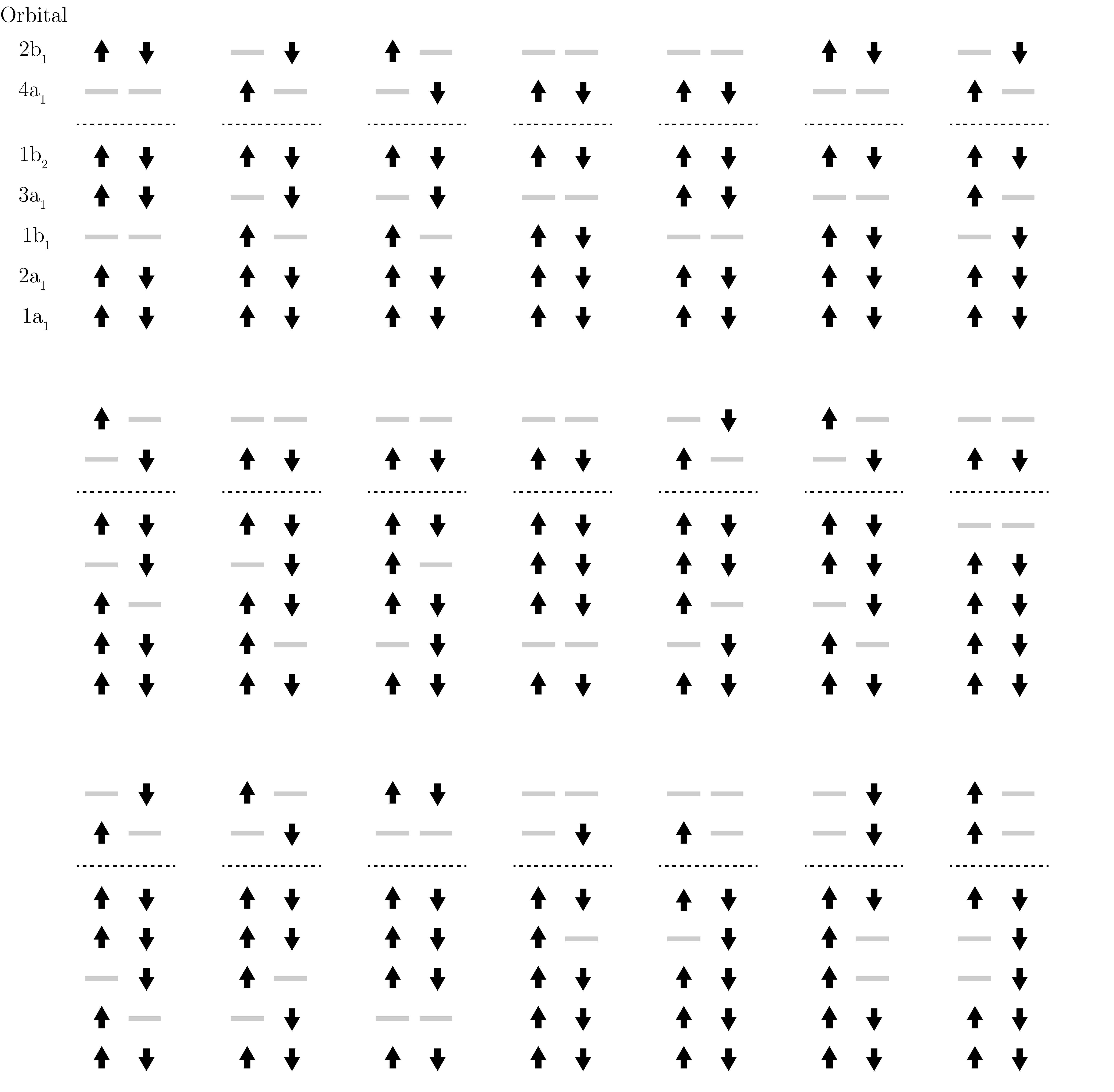}
    \caption{%
   Most significant interaction terms beyond Hartree-Fock.
    \label{fig:allInteractions}
    }
\end{figure*}

Figure~\ref{tab:circ_metric}b shows the  quantum resource requirements for each UCC ansatz circuit optimized for the trapped-ion QC.
Relevant resource metrics include the number of qubits and the number of entangling gates.
We also tabulate the ground state energy from our \textit{in-silico} VQE simulation, as up to 21 terms are added to the ansatz beyond the HF calculations (see Extended Data Fig.~\ref{fig:allInteractions}).
We see that the estimate of the ground state energy approaches the FCI value as more terms are added, reaching the FCI value within chemical accuracy once 17 or more terms are included in the ansatz.
Inspecting the 21 most significant determinants in the FCI energy calculation, we observe that (1) the inner-most MO $1a_1$ (see Figure~\ref{tab:circ_metric}a) is always filled and therefore can be ignored, and (2) the $1b_2$ MO participates only once as a bosonic excitation.
Ignoring $1a_1$ and $1b_2$ can reduce the qubit requirement without sacrificing much in absolute accuracy: the reduced Hamiltonian reaches within 2.1~mHa of the FCI ground state at HF+16 terms using 10 qubits and 140 entangling gates.
Chemical accuracy for the full Hamiltonian is achieved at HF+17 terms with 11 qubits and 143 entangling gates, of which 89 are \cnotgate~gates and 54 are small-angle $\xxgate(\theta)$~gates that feature higher fidelity.
These resource requirements are realistically within the near-term performance targets of a NISQ computer based on trapped ions.

\begin{figure*}[t!]
\centering
\includegraphics[width=\textwidth]{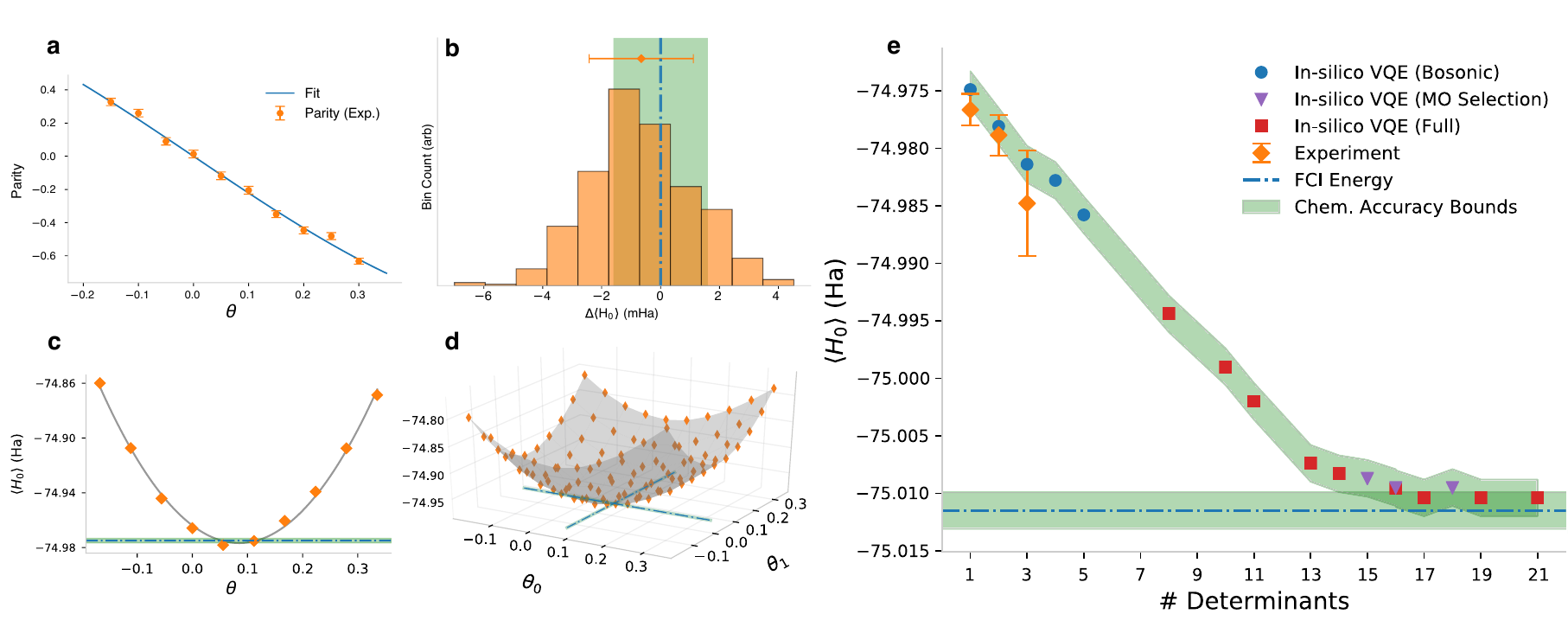}
\caption{
\textbf{Experimental Results.} 
\textbf{a,} Calibration curve for the small-angle $\xxgate(\theta)$~gate.
\textbf{b,} Bootstrap distribution of the HF+2 energy with mean and $1\sigma$ uncertainty indicated by the orange diamond  \textbf{b}, and 
experimentally determined energy surfaces for HF+1 \textbf{c} and HF+2 \textbf{d}. Each data point (orange diamond) represents an average of $\gtrsim 1000$ experimental runs, with the blue dash-dotted lines indicating \textit{in-silico} results.
\textbf{e}, Comparison of ground-state energy estimates as additional interactions are included in the UCC ansatz state (labeled HF+$N$, for $N$ significant determinants).
The orange diamonds indicate experimental results, with error bars indicating $1\sigma$ uncertainty from the bootstrap distribution.
The remaining points are from the \textit{in-silico} VQE simulation as detailed in Figure~\ref{tab:circ_metric}, and show how the ansatz states converge to the full configuration-interaction ground state, indicated by the dot-dashed blue line.
Chemical accuracy about the numeric results is indicated by the shaded green region in all figures.
}
\label{fig:exp}
\end{figure*}

\section{Circuit optimization and co-design}

We have implemented a number of circuit optimization techniques that take advantage of the unique features available in the IonQ trapped-ion QC, but are generic in the sense that they are applicable to any target molecule to be simulated.
The strategies described here are executed by a full-stack, modularized software toolchain, which automatically produces optimized circuits\cite{ar:opt} for generating the ansatz state of a molecular system.

Given a general UCC ansatz state, interaction terms take the form of a two-electron interaction $\theta_{pqrs}c_p^\dagger c_q^\dagger c_r c_s$.
Since the indices $p,q,r,s$ vary over the complete set of molecular states (which are represented by different qubits), implementing this interaction requires entangling gates between arbitrary pairs of qubits in the system.
The all-to-all connectivity of trapped-ion QCs makes this a native operation, eliminating the overhead incurred by repeated \swapgate~gates to reorder qubits before an entangling operation can be applied between nearest-neighbor qubits.
Given that the infidelity of these \swapgate~operations can dominate the quality of complicated computations, eliminating them from the optimized circuit dramatically increases the accuracy of the VQE result.
This circuit optimization is a direct result of co-design for a particular hardware advantage. 

Another general optimization strategy is to represent the bosonic excitations, where two electrons remain paired, as a single creation or annihilation operator. It is convenient to expand the SO label for the operator $c_p$ to $c_{k\alpha}$, where $k$ and $\alpha$ denote the MO and spin label, respectively. Then the bosonic operators are $d^\dagger_k = c_{k\alpha}^\dagger c_{k\beta}^\dagger$ and $d_j = c_{j\alpha} c_{j\beta}$ ($\alpha {\neq} \beta$), which can be directly translated to the Pauli raising/lowering operators $\sigma^j_{\pm}$ on qubit $j$. 
The UCC operator corresponding to the bosonic excitation simplifies to $\exp[\theta_{jk}\sigma^j_{+} \sigma^k_{-} - h.c.]$,
and a pair of arbitrary-angle $\xxgate(\theta)$~gates is sufficient to implement this interaction (see Fig.~\ref{tab:circ_metric}c). 
Thus ansatz states containing only bosonic excitations can be implemented very efficiently on our QC.

For all remaining terms, we must implement the two-electron interaction via the JW transformation.
Each of these terms looks like $\hat{V} = \exp[\theta_{pqrs}\sigma^p_{+}\sigma^q_{+} \sigma^r_{-}\sigma^s_{-} \bigotimes_k \sigma^k_z - h.c.]$, where the product $\bigotimes_k\sigma^k_z$ denotes the adequate JW string to reflect fermionic symmetries. A JW string with $m$ $\sigma_z$ gates converts to $m$ \cnotgate~gates on either side of the subcircuit that would otherwise implement $\hat{V}$. Properly ordering these terms in the entire circuit can eliminate most of the \cnotgate~gates so they represent a relatively low overhead as a function of terms in the UCC ansatz\cite{HastingsQIC2015}. The main portion of the quantum circuit is an implementation of a linear combination of eight terms, each containing a product of four $\sigma_x$ and $\sigma_y$ operators (with odd number of $\sigma_x$ in each term). By optimizing the order of these eight operators and taking advantage of the all-to-all connectivity, we can implement this circuit with 13 \cnotgate~gates (see Fig.~\ref{tab:circ_metric}d). 
When we concatenate several of these terms, some \cnotgate~gates at the ends, including those that arise from a JW string, may cancel out.

Most ansatz states have both bosonic and non-bosonic excitation terms. For these situations, we start with the reduced representation where each qubit describes one MO, and run the quantum circuit that corresponds to all bosonic excitation terms first. Then, additional qubits (all prepared in the $|0\rangle$ state) are introduced, and each are entangled with a qubit representing a MO using a \cnotgate~gate. Each entangled pair can now represent the two SOs corresponding to the MO (Extended Data Fig.~\ref{fig:circuits}e).

One last optimization takes advantage of the asymmetric SPAM error observed in our system. Normally, we encode a filled orbital (MO or SO) with $|1\rangle$ and an empty orbital with $|0\rangle$, but in a molecule with mostly closed molecular shells like ${\rm H}_2 {\rm O}$, the filled orbitals in the HF-ground state remain mostly filled in the FCI ground state as well. Since our SPAM error is more than a factor of two smaller for $|0\rangle$ compared to $|1\rangle$, we  encode the filled orbitals as $|0\rangle$ to reduce the systematic shift associated with readout from the $|1\rangle$ state.
This encoding has the ancillary benefit of requiring fewer single-qubit gates to initialize the circuit, but the advantage diminishes as measurement errors are suppressed or become more symmetric.

Combining these strategies, we achieve the quantum circuits for preparing the ansatz state with total entangling gate counts shown in Fig.~\ref{tab:circ_metric}b. These methods represent a fully general, scalable, and near-optimal framework that can be applied to simulating any physical systems using VQE with a UCC ansatz.

\section{Results}

Using our trapped-ion QC, we compute the first three bosonic excitation terms of the VQE ansatz for the ${\rm H}_2 {\rm O}$ molecule (Extended Data Fig.~\ref{fig:circuits}).
In order to estimate the minimum energy for each circuit within chemical accuracy (fractional uncertainty of ${\sim} \num{e-5}$), all systematic errors in our QC must be carefully characterized and controlled.
The intrinsic decoherence of a $^{171}\textrm{Yb}^+$ trapped-ion qubit is negligible over the timescale of our computation\cite{Yb:2007}, so the dominant errors arise in calibrating the angle of the $\xxgate(\theta)$~gate and correcting for the systematic SPAM error of our ion chain.
We accurately calibrate the angle $\theta$ using a circuit similar to that shown in Fig.~\ref{tab:circ_metric}c, where the parity varies as $\sin(2\theta)$.
Fitting the parity to this functional form (Fig.~\ref{fig:exp}a) compensates for non-linearities in the AOM and enables easy interpolation for arbitrary gate angles.
Uncertainty in the SPAM correction can be made arbitrarily small given sufficient measurement statistics.
Gate fidelity will begin to dominate as the computation length increases, but for the circuits experimentally demonstrated here we are not limited by this error and found no benefit to error mitigation techniques like Richardson extrapolation\cite{ar:RE}.

\begin{figure}
    \centering
    \includegraphics[width=\columnwidth]{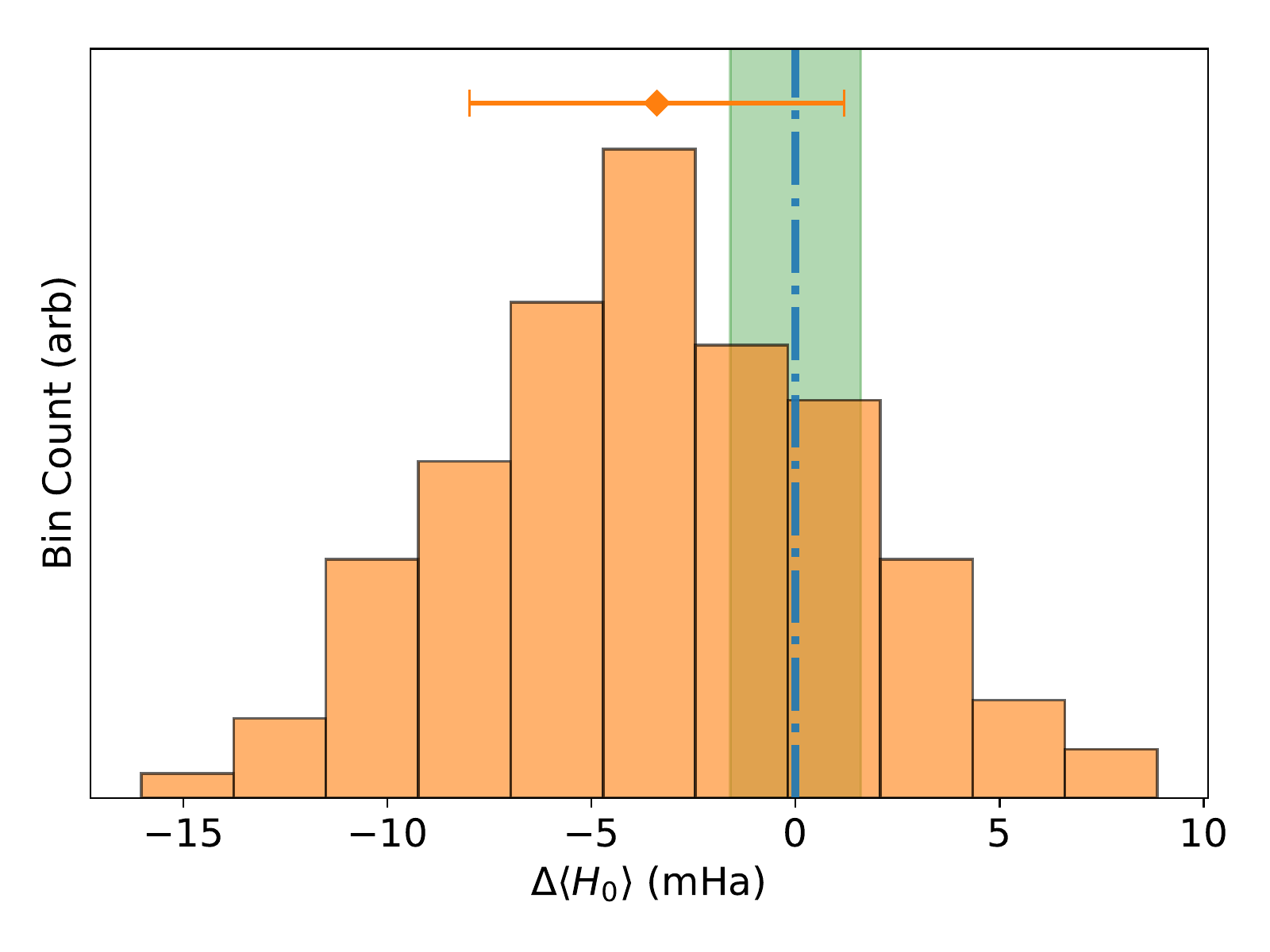}
    \caption{Bootstrap distribution for the experimentally computed energy in HF+3, compared to the \textit{in-silico} result (blue dash-dotted line) and bounds of chemical accuracy (green shaded region). We determine $\langle H_0\rangle = \num{-74.985(5)}$~Ha, indicated by the orange diamond. Error bars signify $1\sigma$ uncertainty. }
    \label{fig:hf3_bs}
\end{figure}

To compute the energy corresponding to a prepared ansatz state, we make a set of projective measurements in bases corresponding to the terms in the Hamiltonian, as previously described.
Once a sufficient number of measurements are made, we use a statistical bootstrapping technique\cite{efron:1994} that accounts for SPAM error to estimate uncertainties from the resulting histograms (Fig.~\ref{fig:exp}b).
Figure~\ref{fig:exp}c-d shows the experimentally determined energy surface for HF+1 and HF+2 as the ansatz parameters $\{\theta_i\}$ are scanned about their optimum values, and the data for HF+3 is shown in the Extended Data Figure~\ref{fig:hf3_bs}. 
The experimentally determined ground-state energies for each of these three ansatz states is $-74.977(1)$~Ha, $-74.979(2)$~Ha, and $-74.985(5)$~Ha, respectively, with parenthetical errors indicating $1\sigma$ uncertainty derived from the bootstrapped distribution.
The dominant experimental uncertainty arises in the SPAM correction, which can be improved with upgrades to the hardware and new tomographic methods\cite{Noek:13,keith2018joint}.
A direct comparison to the \textit{in-silico} VQE simulation can be found in  Fig.~\ref{fig:exp}e.
The match to theory is very good---both the absolute accuracy and precision are comparable to the chemical accuracy.
This is critical to achieve quantum computational results that provide predictability for VQE-type optimization algorithms.
Given the circuit requirements outlined in Fig.~\ref{tab:circ_metric}b and the gate fidelity achieved, we believe that executing a full simulation towards the FCI energy is within reach using current trapped-ion quantum computer technology.

\section{Summary and outlook}
Dramatic improvements must be made to both QC hardware and techniques to efficiently use the available quantum resources in order to perform meaningful quantum computations on a NISQ device.
The work presented here is a framework for end-to-end optimization that maps useful problems in quantum chemistry to a trapped-ion QC, fully leveraging the hardware-specific advantages.
This framework yields near-optimal quantum circuits that can be run on existing NISQ hardware.
To verify the performance of both the hardware and the optimization procedure, we compute the post-Hartree-Fock ground state energy of $\rm H_2\rm O$ on a trapped-ion QC.
Without any error mitigation, the experimental results for the first three correction terms are in excellent agreement with the theory at the level of chemical accuracy, both in the predicted values and their precision.

While these results are specific to a particular quantum chemistry problem and the trapped-ion QC hardware, the computational methodology we develop is completely general to simulating quantum systems. We anticipate that similar advances can be applied to other optimization problems that work on variational methods, such as the quantum approximate optimization algorithm\cite{ar:FGS} and various quantum machine learning applications\cite{Gaoeaat9004,2018arXiv181208862Z}.
Increased attention to co-design principles like those demonstrated here will be necessary to push the boundary of possibility in near-term quantum computation.


\section*{Acknowledgements}
The authors would like to thank David Moehring for his guidance in the construction of this apparatus, and the EURIQA team at the University of Maryland and Duke University for sharing their designs and for useful conversations.

\section*{Author Contributions}
Experimental data collected and analyzed by N.C.P. and JS.C.; Y.N., C.D., D.M. K.B. and J.K. performed the circuit design and Y.N. performed \textit{in-silico} VQE simulation; the apparatus was designed and built by  K.W., J.M.A., K.M.B., JS.C., M.C., S.D., K.M.H., J.M., N.C.P., J.D.WC., S.M.K, S.A, J.A., P.S., M.W., A.M.D., A.B., V.C., M.K., C.C., C.M. and J.K.; Y.N., N.C.P, JS.C. and J.K. prepared the manuscript, with input from all authors.

\section*{Competing Interests}
The authors declare no competing interests.
\section*{Correspondence} 
Correspondence and requests for materials should be addressed to Yunseong Nam~(email: nam@ionq.co) or Jungsang Kim (kim@ionq.co).

\bibliography{reference}





\section*{Methods}
\label{sec:Methods}

\subsection{SPAM characterization and correction.}
\label{sec:spam}

We monitor the SPAM error during the computation by interleaving experiments that prepare the all-bright {$|11 \cdots 11\rangle$} and all-dark {$|00 \cdots 00\rangle$} states of the ion chain.
During detection, we relax the axial confinement such that the crosstalk error  between adjacent ions is below the intrinsic dark count error of our photon detectors.
This enables a fast measurement of the entire SPAM matrix, because it decomposes into a Kronecker product of SPAM matrices on the individual ions.
The average single-ion SPAM errors, along with an example SPAM matrix of a three-ion chain, are summarized in Fig.~\ref{fig:apparatus}b, which are independently verified via single-qubit randomized benchmarking\cite{Knill:2008}.
The measured state vectors are then corrected for SPAM error via matrix inversion before computing expectation values of the circuit Hamiltonian~\cite{ShenNJP2012,ballance:2016}.
Note that for the QC performing computations here, the SPAM error is dominated by detection fidelity\cite{acton2006near}, which can be improved significantly by upgrading the photon collection efficiency of the hardware.

\subsection{Single-qubit operation.}
\label{sec:1qubit}

\begin{figure}
    \centering
    \includegraphics[width=\columnwidth]{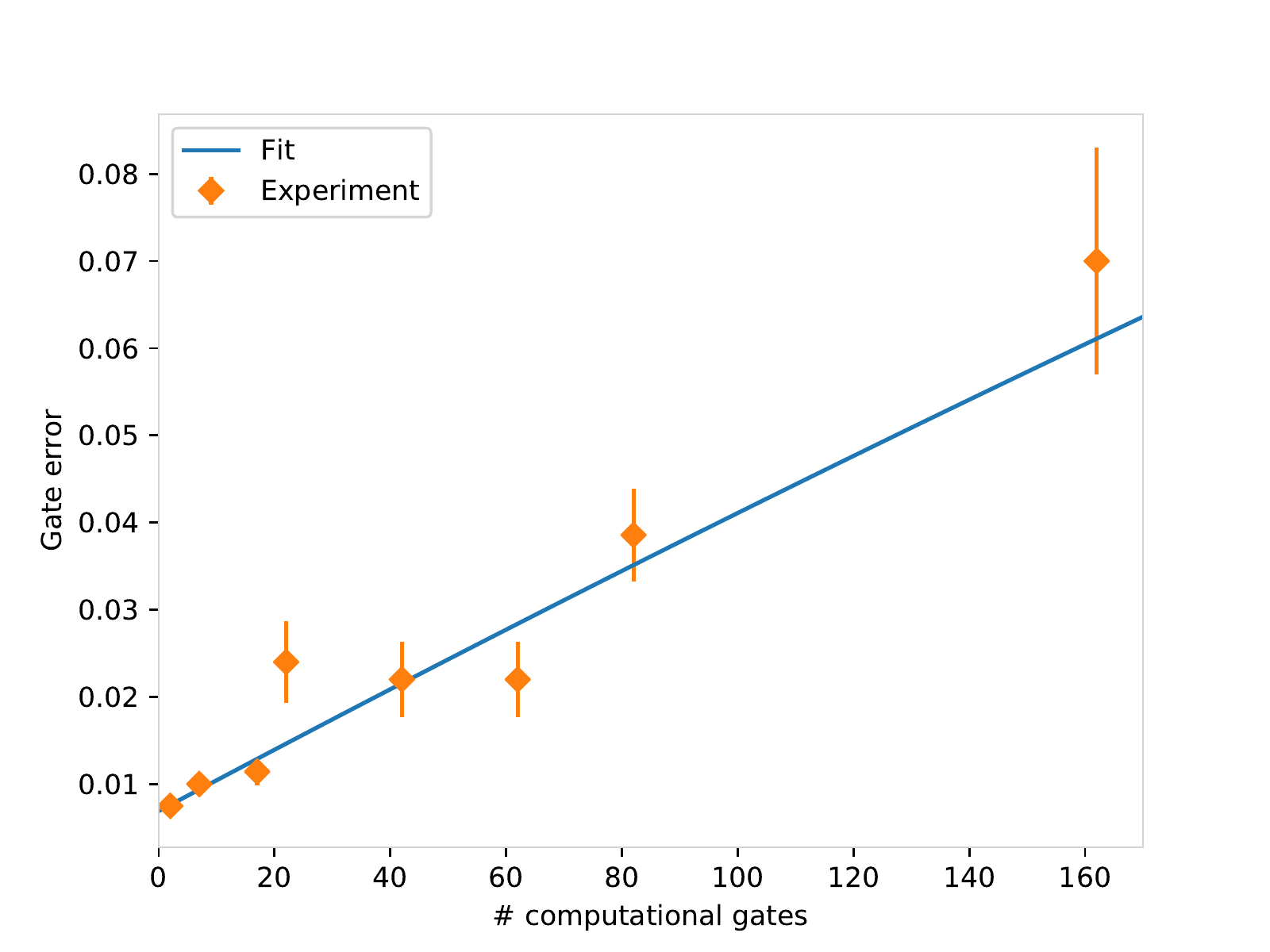}
    \caption{%
    Typical randomized benchmarkting data for single-qubit gates, using an SK1 composite pulse sequence. The extracted fidelity is $99.964(5)\%$. 
    \label{fig:rb}
    }
\end{figure}

Single-qubit gates are implemented via Rabi oscillations between $|0\rangle$ and $|1\rangle$, using an SK1 composite pulse sequence\cite{Brown:2004}.
The transition is driven by a two-photon Raman process from two separate beams generated by a pulsed \SI{355}{\nano\meter} laser, where one beam is tightly focused onto the ion so it can be manipulated independently from the rest of the chain.
We prepare arbitrary single-qubit states on the Bloch sphere by controlling the phase and pulse area delivered by the individual addressing beam. 
Conventional randomized benchmarking\cite{Knill:2008} and gate set tomography\cite{blume2017demonstration} techniques are used to characterize our single-qubit gates, and we observe fidelities ${\gtrsim} 99.9\%$ with good repeatability. A typical randomized benchmarking result is depicted in the Extended Data Fig.~\ref{fig:rb}. 
During the computation, we interleave  experiments to calibrate the laser power based on measured Rabi rates.
Single-qubit gates are also used to measure SPAM error in the $|1\rangle$ state.
Although we do not characterize the single-qubit gate fidelity during the computation, we are able to bound it ${\gtrsim}99\%$ by monitoring the SPAM error.

\subsection{Two-qubit operation.}
\label{sec:2qubit}
Two-qubit entangling gates are implemented via the {M{\o}lmer}-{S{\o}renson} interaction, where an amplitude-modulated laser pulse, composed of non-copropagating beams, achieves full spin-motion decoupling at the end of the gate\cite{ZhuPRL2006, ZhuEL2006,Choi:XX,DebnathNature2016,ar:ATA}.
One of the laser beams uniformly illuminates the ions, while the other individually addresses the two particular ions involved in the gate. By varying the pulse area through the laser intensities, we control the geometric phase $\theta$ of the interaction, defined in terms of the Pauli-$X$ matrices $\sigma_x^i$ on the $i$-th ion as
\begin{align}
   \xxgate_{ij}(\theta) \equiv e^{-i\theta\sigma_x^i\sigma_x^j/2}\,, \label{eq:xx-U}
\end{align}
where $\theta = \pi /2$ corresponds to a maximally  entangling gate.
We calibrate the small-angle $\xxgate$~gate $(\theta \ll \pi /2)$ with the circuit presented in Extended Data Fig.~\ref{fig:circuits}a, projecting to the $\sigma_x$ basis before measurement.
An overall amplitude scale factor $g$, applied to the acousto-optic modulator (AOM) controlling the individual addressing beams, is scanned to vary the geometric phase $\theta \propto g^2 \equiv \Theta$, with $g^2 < 0$ realized by shifting the rf phase of the AOM thereby inverting $\Theta \rightarrow -\Theta$.
The parity $\Pi_{ij}$ of the output state, defined in terms of the observed two-qubit state probability $P_{ij}$ as
\begin{equation}
    \Pi_{ij} = P_{ij}\left(|00\rangle\right) + P_{ij}\left(|11\rangle\right) - P_{ij}\left(|10\rangle\right) - P_{ij}\left(|01\rangle\right)\text{,} 
\end{equation}
is fit to $\Pi_{ij} = \sin(2k\Theta)$, and the factor $k$ is applied to the AOM amplitude scale factor $g$ to  properly calibrate arbitrary-angle $\xxgate(\theta)$~gates.
An example calibration curve is shown in Fig.~\ref{fig:exp}a.
Small-angle two qubit gates have high fidelity, which we estimate to be $\gtrsim 99.6\%$ for the $\theta = \pi /100$ case (Fig.~\ref{fig:apparatus}d), which is likely explained by the smaller absolute geometric phase errors and weaker light shifts.
A method for rigorous characterization of an arbitrary-angle \xxgate~gate fidelity remains an open question for future study.

\subsection{Water molecule.}
We use the nuclear configuration of $\rm H_2\rm O$ where the ${\rm O}{-}{\rm H}$ bond length is fixed to ${\sim} 1.8a_0$ and the angle between the two ${\rm O}{-}{\rm H}$ bonds to be ${\sim} 105 \degree$, where $a_0$ is the Bohr radius.

\subsection{Molecular modeling with qubits.}
\label{sec:Overview}

We start with the second-quantized Hamiltonian of the form
\begin{equation} 
\label{eq:Phys_H}
\hat{H} = \sum_{p,q}{t_{pq} a_p^\dagger a_q} + \sum_{p,q,r,s}{t_{pqrs} a_p^\dagger a_q^\dagger a_r a_s},
\end{equation}
where $a_p^{\dagger} (a_p)$ is the creation (annihilation) operator for an atomic orbital (AO) with a given spin. We evaluate $t_{pq} = {\int}\hspace{-.2em}{\int} \chi^{*(1)}_p \hat{H}_1 \chi^{(2)}_q d\vec{r}^{(1)}d\vec{r}^{(2)}$ and $t_{pqrs} = {\int}\hspace{-.2em}{\int} \chi^{*(1)}_p \chi^{*(2)}_q \hat{H}_2 \chi^{(2)}_r \chi^{(1)}_s d\vec{r}^{(1)}d\vec{r}^{(2)}$ using the minimal, STO-3G basis \cite{ar:STO3G}, where $\chi^{(i)}_p$ is the $p$-th AO for $i$-th electron and $\hat{H}_j$ is the $j$-body electronic Hamiltonian. We then employ the HF method and solve the Roothan equation using the self-consistent approach\cite{bk:MC} to obtain molecular orbitals (MOs) as a linear combination of AOs. Each MO consists of two spin orbitals (SOs) corresponding to each spin state of the electron (up and down). The resulting Hamiltonian in SOs is given in Eq.~(\ref{eq:hamiltonian}) of the main text.

We then apply Jordan-Wigner (JW) transform\cite{ar:JW}
\begin{equation} 
\label{eq:JW}
c_j^\dagger  = 1^{\otimes j-1} \otimes \sigma^j_+ \otimes \sigma_z^{\otimes N - j}, \quad c_j  = 1^{\otimes j-1} \otimes \sigma^j_- \otimes \sigma_z^{\otimes N - j}
\end{equation}
to transform our reduced, physical, SO-basis Hamiltonian to a qubit Hamiltonian. Additionally, we choose the unitary coupled-cluster (UCC) as our ansatz\cite{ar:UCC1,ar:UCC2} and apply the same transformation. Specifically, for the circuit-level implementation of the JW-transformed ansatz circuit $\exp(-iH_{c.c.})$, the effective Hamiltonian, coupled-cluster (CC) operator $H_{c.c.}$ is given by $H_{c.c.} = i(T-T^\dagger)$, where $T$ is the excitation operator. We use $n$-th order product-formula (PF) algorithms to further decompose the CC operator, according to
\begin{equation}
\exp\biggl(-i\sum_{j=1}^{L}\theta_jH_{j}\biggr) \approx [S_{n}(\lambda)]^{r},
\end{equation}
where $H_{c.c.} =  \sum_j \theta_j H_j$, $\lambda := 1/r$, and
\begin{align}
S_{1}(\lambda)&:=\prod_{j=1}^{L}\exp(-i\theta_jH_{j}\lambda), \nonumber \\
S_{2}(\lambda)&:=\prod_{j=1}^{L}\exp(-i\theta_jH_{j}\lambda/2)\prod_{j=L}^{1}\exp(-i\theta_jH_{j}\lambda/2), \nonumber \\
S_{2k}(\lambda)&:=[S_{2k-2}(p_{k}\lambda)]^{2}S_{2k-2}((1-4p_{k})\lambda)[S_{2k-2}(p_{k}\lambda)]^{2},
\label{eq:recursive_def}
\end{align}
with $p_{k}:=1/(4-4^{1/(2k-1)})$ for $k>1$ \cite{ar:Suzuki}. We choose the PF algorithm among many available methods, as the PF resulted in the smallest algorithmic error for certain Hamiltonian simulations while requiring the least quantum resources\cite{ar:PNAS}.
We find it sufficient to use the first order Trotter formula ($n=1$) with a single PF stage ($r=1$), as the resulting ansatz leads to the FCI ground state energy within chemical accuracy. 

\subsection{Expectation values of the Hamiltonian.}
Once the ansatz state is prepared, the expected energy of the state is evaluated by projecting the state onto a set of different multi-qubit Pauli bases, obtained from the JW transformation of the SO Hamiltonian in Eq.~(\ref{eq:hamiltonian}). Specifically, a set of basis-transformation operations, depicted in Extended~Data~Fig.~\ref{fig:circuits}d, are applied prior to measurement in the computational basis.
The expectation of each Pauli string can be computed at the desired level of accuracy by sufficient statistical sampling.
Due to the coefficients in the Hamiltonian, some terms contribute more than others to the final computed energy.
We minimized the number of basis transformations required to evaluate the entire Hamiltonian by carefully partitioning the Pauli strings, grouping those that can be measured with a single, representative measurement circuit.

\subsection{Circuit for bosonic excitation terms.}
\label{sec:CircBosonic}

For the cases where the pair of electrons from one MO are excited together to another MO, we can represent this process as a ``bosonic'' excitation. It is convenient to expand the SO index to reflect both MO ($j$, $k$, etc.) and spin ($\alpha$ or $\beta$) indices, {\it e.g.},  $c_p \equiv c_{k\alpha}$. Then the bosonic creation and annihilation operators are $d^\dagger_k = c_{k\alpha}^\dagger c_{k\beta}^\dagger$ or $d_j = c_{j\alpha} c_{j\beta}$ ($\alpha \neq \beta$).
 In these bosonic excitations, we can use one creation (annihilation) operator to represent the occupation of both electrons in that orbital. Since these creation and annihilation operators now correspond to a pair of electrons, the anti-commutation relations that apply to fermionic particles no longer hold---thus the term ``bosonic excitation''---and the $\sigma^{i}_z$ strings in the JW transformation are not necessary. For these bosonic excitation terms, the operator $d^\dagger_k$ ($d_k$) can be replaced by the qubit operator $\sigma^k_+$ ($\sigma^k_-$), and the corresponding UCC operator reduces to $\exp[\theta_{kj}\sigma^k_+ \sigma^j_- - h.c.]$. The implementation of this term in a quantum circuit requires two \xxgate~gates between the two qubits (\textit{k} and \textit{j}) with an angle of $\theta_{kj}$, denoted as $\xxgate_{kj}(\theta_{kj})$, as shown in Fig.~\ref{tab:circ_metric}c.

\subsection{Circuit for non-bosonic excitation terms.}
\label{sec:CircDesign}

Several steps are taken to minimize the number of entangling gates in implementing the quantum circuit for generic two-electron interaction terms.

1. First, we need to allocate a qubit to represent each SO considered in the molecule. By putting frequently-interacting SOs closer to each other, we can reduce the product of $\sigma_z$ operators introduced in the JW transformation (called JW strings). A simple greedy approach is used to identify the most frequent interaction between different SOs, and the identified SO pairs are mapped to nearest-possible qubits until no further mapping may be made. 

2. An adequate ordering of the individual terms in the JW-transformed $H_{c.c}$ in the Trotterized ansatz circuit can lead to dramatic reduction in the JW strings between the terms~\cite{HastingsQIC2015}. We order the terms in the ansatz states such that the adjacent terms have maximal overlap in non-identity elements in the Pauli product, while making sure that at least one overlap is non-$\sigma_z$. This maximizes the cancellation of the JW strings between the adjacent terms, simplifying the resulting circuit.

3. A two-electron excitation term that consists of  ${\rm Im}(c^{\dagger}_p c^{\dagger}_q c_r c_s)$ translates to the product operator ${\rm Im}(\sigma_+^p\sigma_+^q\sigma_-^r\sigma_-^s \bigotimes_{\nu}  \sigma_z^\nu)$ by the JW transformation, where $\sigma_\pm = (\sigma_x \mp i \sigma_y)/2$ and the $\sigma_z$ product is the JW string.
The corresponding unitary operator takes the form $\exp[-i\theta(\bigotimes_\nu \sigma_i^{\nu})/2]$, 
where $\sigma_i^{\nu} \in \{\mathbbm{1},\sigma_x,\sigma_y,\sigma_z\}$ is the Pauli operator for $\nu$-th qubit. We implement the circuit as shown in \eq{ZZZZ}a below, instead of a nearest-neighbor-inspired, ``staircase'' construction (up to \swapgate~gates) such as in \eq{ZZZZ}b. Here, we assumed that all $\sigma_i^\nu = \sigma_z$ as a concrete example. The circuit implementation of the term $\exp[-i\theta\bigotimes_{\nu=0}^{N-1} \sigma_i^{\nu}/2] $ results in $2(m-1)$ $\cnotgate$ gates, where $m$ is the number of elements $\sigma_i$ that are not identity, and the common target qubit for the  $\cnotgate$ gates can be chosen arbitrarily among the $m$ qubits.
\vspace{-0.3em}
\begin{align}
\hspace{-0.4em}
{\bf a.} \,\,
\scalebox{0.7}{\mbox{\Qcircuit @C=0.5em @R=0.02em @!R{
&\qw &\ctrl{3}  &\qw      &\qw      &\qw                            &\qw     &\qw     &\ctrl{3} &\qw &\qw\\
&\qw &\qw      &\ctrl{2}  &\qw      &\qw                            &\qw     &\ctrl{2} &\qw     &\qw &\qw\\
&\qw &\qw      &\qw       &\ctrl{1} &\qw                            &\ctrl{1} &\qw     &\qw     &\qw &\qw\\
&\qw &\targ    &\targ     &\targ     &\gate{\rzcirc(\theta)} &\targ    &\targ   &\targ    &\qw &\qw }}}
\qquad\;\;
{\bf b.} \,\,
\scalebox{0.7}{\mbox{\Qcircuit @C=0.5em @R=0.02em @!R{
&\qw &\ctrl{1} &\qw     &\qw     &\qw                            &\qw     &\qw      &\ctrl{1} &\qw &\qw\\
&\qw &\targ    &\ctrl{1} &\qw     &\qw                            &\qw     &\ctrl{1} &\targ   &\qw &\qw\\
&\qw &\qw     &\targ    &\ctrl{1} &\qw                            &\ctrl{1} &\targ   &\qw     &\qw &\qw\\
&\qw &\qw     &\qw      &\targ    &\gate{\rzcirc(\theta)} &\targ    &\qw    &\qw     &\qw &\qw }}}
\label{eq:ZZZZ}
\end{align}

 To optimize the resulting quantum circuit, we start by choosing the common target to be one of the four qubits to which either $\sigma_x$ or $\sigma_y$ is applied. Here, as a concrete example, we choose the last such qubit as the common target. Once the JW strings are expressed by the \cnotgate~gates, the rest of the circuit consists of eight terms that can be ordered as $\sigma_x\sigma_x\sigma_x\sigma_y$, $\sigma_x\sigma_x\sigma_y\sigma_x$, $\sigma_y\sigma_x\sigma_y\sigma_y$, $\sigma_y\sigma_x\sigma_x\sigma_x$, $\sigma_y\sigma_y\sigma_x\sigma_y$, $\sigma_y\sigma_y\sigma_y\sigma_x$, $\sigma_x\sigma_y\sigma_y\sigma_y$, and $\sigma_x\sigma_y\sigma_x\sigma_x$.  This ordering, while not a unique example, allows us to better optimize the quantum circuit. First, we perform basis transformation $\sigma_i \mapsto \sigma_z$ by applying $\hgate$ or $\sgate^\dagger \hgate$ for $i = x$ or $y$, respectively. Then, the resulting term always looks like a product of four $\sigma_z$ gates, and the circuit shown in \eq{ZZZZ}a can be used to implement each term. The particular order has exactly two basis changes ($\sigma_x \leftrightarrow \sigma_y$) between adjacent terms resulting in the cancellation of two pairs of \cnotgate~gates between them, leaving only two \cnotgate~gates between each term.
This cancellation is not applicable for the staircase implementation shown in \eq{ZZZZ}b. 

4. The two \cnotgate~gates between the two terms discussed above can be further simplified, and reduced to a single \cnotgate~gate using  
the following circuit identity:
\begin{align}
\mbox{\Qcircuit @C=0.5em @R=0.7em {
&\qw &\ctrl{1} &\gate{\hcirc} &\ctrl{1} &\qw &\qw\\
&\qw &\targ    &\qw                &\targ    &\qw &\qw }}
\hspace{2mm}\raisebox{-3.3mm}{$=$}\hspace{3mm}
\mbox{\Qcircuit @C=0.5em @R=0.4em {
&\gate{\scirc}  &\targ     &\gate{\scirc^\dagger}   &\qw \\
&\gate{\hcirc}  &\ctrl{-1} &\gate{\scirc}           &\qw  }}\,\hspace{2mm}\raisebox{-3.3mm}{.}\hspace{3mm}
\end{align}
Noting that any $z$-rotations, including $\sgate$ and $\sgate^\dagger$, commute with the control of a $\cnotgate$ gate, and $\hgate \sgate \hgate = \sgate^\dagger \hgate \sgate^\dagger$ and $\hgate \sgate^\dagger \hgate = \sgate \hgate \sgate$ up to a global phase, we can further simplify the single-qubit gates in the circuit by commuting the $\sgate$, $\sgate^\dagger$, and $\hgate$ gates to either end of the circuit through the $\cnotgate$ gates. The resulting circuit to simulate any two-electron term reduces to Fig.~\ref{tab:circ_metric}d, and consists of 13 \cnotgate~gates, eight small-angle rotations about $z$-axis, and a few $\hgate$, $\sgate$, and $\sgate^\dagger$ gates to account for the relevant basis rotations.

\subsection{Circuit efficiency.}

For a two-electron interaction term over four qubits, we start with eight instances of the circuit shown in \eq{ZZZZ} containing 48 $\cnotgate$ gates. 
Our final template circuit in Fig.~\ref{tab:circ_metric}d, which contains 13 entangling gates, is likely optimal. 
At least one multi-qubit gate needs to be expended to transform from one Pauli basis to another, and our design requires exactly one two-qubit gate for such transformation. 
If a quantum circuit is written with  $\cnotgate$  as the only available multi-qubit gate, it takes at least three $\cnotgate$~gates to compute or uncompute a Pauli product of length four, such as $\sigma_x\sigma_x\sigma_x\sigma_y$.
Given each two-electron excitation operator in the PF-based implementation includes eight such Pauli products, the 13 $\cnotgate$~gates in Fig.~\ref{tab:circ_metric}d is likely a minimum.  This circuit optimization is completely general for a two-electron interaction term applied to any molecule using any basis set and any order of PF (see \eq{recursive_def}).
 
\subsection{Higher fidelity two-qubit gates.}
\label{sec:CNOTtoXX}

The number of entangling gates is not the sole factor that determines the quality of a quantum computation: it also depends on the type of entangling gate used. Specifically, a small-angle \xxgate~gate performs better in our trapped-ion QC than a $\cnotgate$ gate, which requires the maximally entangling $\xxgate(\pi/2)$ gate. Therefore, it is advantageous to convert $\cnotgate$ gates to small-angle $\xxgate$ gates wherever possible.

Shown below is a circuit replacement rule that may be used to convert $\cnotgate$ gates to $\xxgate(\theta)$ gates.  The rule can be applied to the target circuit in Fig.~\ref{tab:circ_metric}d to replace four \cnotgate~gates with four \xxgate~gates. 
Assuming the infidelity associated with a small-angle $\xxgate$ gate is much lower than that of a $\cnotgate$~gate, applying this procedure replaces four out of 13 $\cnotgate$ gates with small-angle $\xxgate$ gates, bringing about a ${\sim}30\%$ reduction in the two-electron interaction infidelity.
\[
\scalebox{0.8}{\mbox{\Qcircuit @C=0.21em @R=0.02em @!R{
& &\qw &\targ     &\gate{\rzcirc(\theta)} &\targ    &\gate{\rzcirc(-\theta)} &\targ    &\qw \\
& &\qw &\ctrl{-1} &\qw            &\qw      &\qw                 &\ctrl{-1}&\qw \\ 
& &\qw &\qw       &\qw            &\ctrl{-2}&\qw                 &\qw      &\qw }}
\hspace{2mm}\raisebox{-5.5mm}{$=$}\hspace{2mm}
\mbox{\Qcircuit @C=0.21em @R=0.49em @!R{
&\qw &\gate{\hcirc} &\multigate{1}{\xxcirc(\theta)} &\gate{\hcirc}  &\targ     &\gate{\hcirc} &\multigate{1}{\xxcirc(-\theta)} &\gate{\hcirc} &\qw\\
&\qw &\gate{\hcirc} &\ghost{\xxcirc(\theta)}        &\qw              &\qw       &\qw           &\ghost{\xxcirc(-\theta)}       &\gate{\hcirc} &\qw\\
&\qw &\qw           &\qw                    &\qw              &\ctrl{-2} &\qw           &\qw                                &\qw &\qw
}}}
\]

\subsection{Combining bosonic and non-bosonic terms.}
\label{sec:SystOpt}

\begin{figure*}[t]
\includegraphics[width=\textwidth]{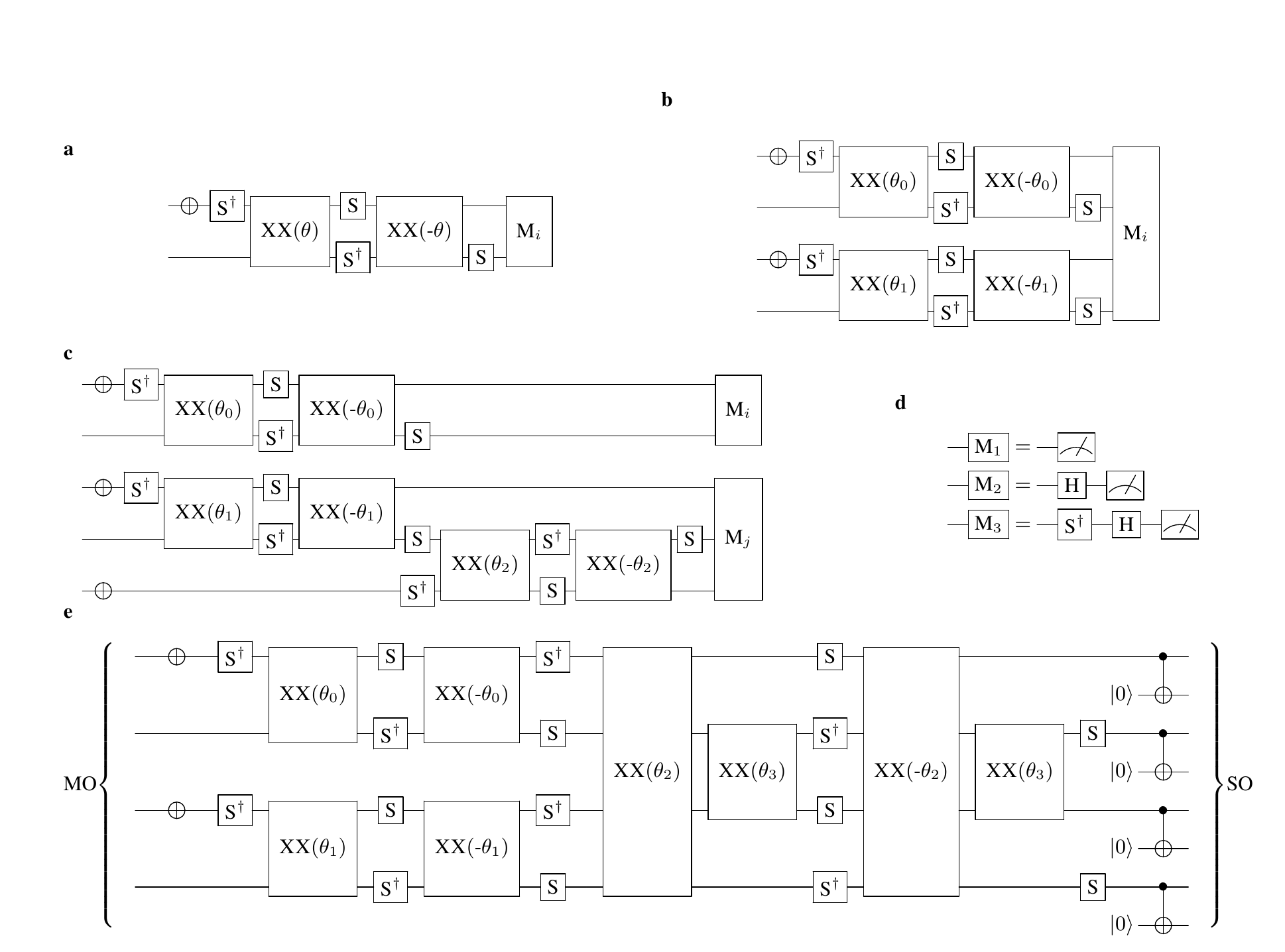}
    \caption{VQE ansatz circuits for \textbf{a,} HF+1, \textbf{b,} HF+2, and \textbf{c,} HF+3. 
    Each circuit ends in a measurement $\textrm M_i$, which is one of three measurement circuits shown in \textbf{d}, to generate expectation values of each Pauli operator.
    For HF+1, shown here in the bosonic representation (see text), the electron-pair excitation from the $1b_1$ MO to the $2b_1$ MO is achieved by applying two $\xxgate(\theta)$ gates between qubits denoting the occupation of electrons in these MOs. 
    The second bosonic excitation added in HF+2 (corresponding to electron pairs from the $3a_1$ MO being excited to the $4a_1$ MO) involves a completely independent set of MOs, so it amounts to parallel execution of the circuit in \textbf{a} with different values of the ansatz parameters $\theta_0$ and $\theta_1$.
For  HF+3, the circuit involves six $\xxgate(\theta)$ gates with three independent parameters.
    All interaction terms are listed in Extended Data Figure~\ref{fig:allInteractions}.
    The \notgate~gates can be commuted through the entire circuit and applied classically after the measurement which, due to asymmetric SPAM error in the $|1\rangle$ state, improves the fidelity of the computation.
    \textbf{e,} A sample circuit mixing bosonic and non-bosonic excitation terms.
    The bosonic excitations are simulated first on a subset of the qubits representing the electron pairs in each MO, then \cnotgate~gates between each MO qubit and an ancilla serve to re-encode the MOs as SOs and the circuit proceeds to execute the remaining non-bosonic interactions.
    For all \xxgate~gates, the two-qubit interaction is applied between the top and bottom qubit lines enclosed by the box.
    \label{fig:circuits}
    }
\end{figure*}

A typical UCC ansatz state contains both bosonic and non-bosonic excitation terms. In this general case, we first assign $N$ qubits to represent the occupation of $N$ MOs with a pair of spin-up and spin-down electrons, and execute the circuit corresponding to all bosonic excitation terms first. Then, we introduce an additional $N$ ancilla qubits all prepared in $|0\rangle$ state. We apply $N$ \cnotgate~gates, each employing a qubit used in simulating the bosonic excitation terms as the control qubit and a fresh ancilla qubit as the target. Each pair can now be used to represent the two SOs corresponding to the MO in executing the remaining non-bosonic excitation terms in the circuit. An example circuit for $N=4$ is shown in Extended Data Figure~\ref{fig:circuits}e.

The savings obtained from employing this hybrid approach in executing the bosonic and non-bosonic excitation terms diminish as the problem size increases, as there are more non-pair excitations than pair excitations in general. Nevertheless, for NISQ devices where every saving matters, we find the reduction to be non-negligible.

\subsection{Selection of spin-orbitals.}
\label{sec:MOSel}

In the simulation for small molecules such as H$_2$O where the solution can be simulated classically, we have full knowledge over which SOs, each mapping to a qubit, participate in our simulations. 
In such cases where the information regarding which SOs are significant in finding the ground state of the molecule, we may drop those SOs that are least significant from our simulation. Such information is likely accessible even in the superclassical regime since, for instance, we may reasonably expect that core orbitals are less likely to participate in the excitation than valence orbitals. 

Restricting the simulation to the most significant SOs results in qubit count savings, at the small cost of leaving out those less significant interactions that should in principle be accounted for in obtaining the ground state energy. In terms of implementation, one may allocate the freed-up quantum resources to simulating interactions between the more significant SOs. This is thus useful whenever the number of qubits is a limiting factor.

\subsection{Bootstrap error analysis.}

In deriving error estimates for the computed energies, we employ the empirical Bootstrap technique\cite{efron:1994}.
All single-shot samples for a particular implementation of the circuit are binned, and a random bootstrap sample $S^*$ of the same size as the original dataset is drawn with replacement from the data.
Similarly, we bin all SPAM characterization data acquired during a particular run, and draw a bootstrap sample from that data as well.
This is repeated five hundred times, and with each bootstrap sample we compute the circuit energy expectation $\langle H_0\rangle$ to build a histogram of possible measurements consistent with the empirical data.
The mean of this distribution is reported as the measured value, and the standard deviation provides a $1\sigma$ error estimate.


\renewcommand{\figurename}{Extended Data Figure}

\end{document}